\documentclass{PoS}

\usepackage{rotating}

\usepackage{ifthen}

\newboolean{inbibliography}

\newboolean{uprightparticles}
\setboolean{uprightparticles}{true} 
\usepackage{xspace} 
\usepackage{upgreek}

\def\lhcb {\mbox{LHCb}\xspace}

\def\spd    {SPD\xspace}
\def\presh  {PS\xspace}
\def\ecal   {ECAL\xspace}
\def\hcal   {HCAL\xspace}
\def\MagUp {\mbox{\em Mag\kern -0.05em Up}\xspace}

\ifthenelse{\boolean{uprightparticles}}%
{

 \def\Ppi         {\ensuremath{\uppi}\xspace}

 \def\PDelta      {\ensuremath{\Delta}\xspace}                 
 \def\PXi      {\ensuremath{\Xi}\xspace}                 
 \def\PLambda      {\ensuremath{\Lambda}\xspace}                 
 \def\PSigma      {\ensuremath{\Sigma}\xspace}                 
 \def\POmega      {\ensuremath{\Omega}\xspace}                 
 \def\PUpsilon      {\ensuremath{\Upsilon}\xspace}                 
 
 \def\PB      {\ensuremath{\mathrm{B}}\xspace}                 
                  
 \def\PD      {\ensuremath{\mathrm{D}}\xspace}

 \def\PK      {\ensuremath{\mathrm{K}}\xspace}

 \def\Pb      {\ensuremath{\mathrm{b}}\xspace}

 \def\Pi      {\ensuremath{\mathrm{i}}\xspace}

}
{

 \def\Ppi         {\ensuremath{\pi}\xspace}

 \mathchardef\PDelta="7101
 \mathchardef\PXi="7104
 \mathchardef\PLambda="7103
 \mathchardef\PSigma="7106
 \mathchardef\POmega="710A
 \mathchardef\PUpsilon="7107
                  
 \def\PB      {\ensuremath{B}\xspace}                 
                  
 \def\PD      {\ensuremath{D}\xspace}

 \def\PK      {\ensuremath{K}\xspace}

 \def\Pb      {\ensuremath{b}\xspace}

 \def\Pi      {\ensuremath{i}\xspace}

}

\makeatletter
\ifcase \@ptsize \relax
  \newcommand{\miniscule}{\@setfontsize\miniscule{4}{5}}
\or
  \newcommand{\miniscule}{\@setfontsize\miniscule{5}{6}}
\or
  \newcommand{\miniscule}{\@setfontsize\miniscule{5}{6}}
\fi
\makeatother

\DeclareRobustCommand{\optbar}[1]{\shortstack{{\miniscule (\rule[.5ex]{1.25em}{.18mm})}
  \\ [-.7ex] $#1$}}

\def\bquark    {{\ensuremath{\Pb}}\xspace}
\def\bquarkbar {{\ensuremath{\overline \bquark}}\xspace}
\def\bbbar     {{\ensuremath{\bquark\bquarkbar}}\xspace}

\def\pion   {{\ensuremath{\Ppi}}\xspace}
\def\piz    {{\ensuremath{\pion^0}}\xspace}

  \def\Kbar    {{\kern 0.2em\overline{\kern -0.2em \PK}{}}\xspace}

\def\KorKbar    {\kern 0.18em\optbar{\kern -0.18em K}{}\xspace}

  \def\Dbar    {{\kern 0.2em\overline{\kern -0.2em \PD}{}}\xspace}

\def\DorDbar    {\kern 0.18em\optbar{\kern -0.18em D}{}\xspace}

\def\Bbar    {{\ensuremath{\kern 0.18em\overline{\kern -0.18em \PB}{}}}\xspace}

\def\BorBbar    {\kern 0.18em\optbar{\kern -0.18em B}{}\xspace}

  \def\Y#1S{\ensuremath{\PUpsilon{(#1S)}}\xspace}

\def\Lbar        {{\ensuremath{\kern 0.1em\overline{\kern -0.1em\PLambda}}}\xspace}
\def\LorLbar    {\kern 0.18em\optbar{\kern -0.18em \PLambda}{}\xspace}

\def\to                 {\ensuremath{\rightarrow}\xspace}

\def\AT#1     {\ensuremath{A_{\mathrm{T}}^{#1}}\xspace}           

\def\C#1      {\ensuremath{\mathcal{C}_{#1}}\xspace}                       
\def\Cp#1     {\ensuremath{\mathcal{C}_{#1}^{'}}\xspace}                    
\def\Ceff#1   {\ensuremath{\mathcal{C}_{#1}^{\mathrm{(eff)}}}\xspace}        
\def\Cpeff#1  {\ensuremath{\mathcal{C}_{#1}^{'\mathrm{(eff)}}}\xspace}       
\def\Ope#1    {\ensuremath{\mathcal{O}_{#1}}\xspace}                       
\def\Opep#1   {\ensuremath{\mathcal{O}_{#1}^{'}}\xspace}                    

\newcommand{\tev}{\ifthenelse{\boolean{inbibliography}}{\ensuremath{~T\kern -0.05em eV}\xspace}{\ensuremath{\mathrm{\,Te\kern -0.1em V}}}\xspace}
\newcommand{\gev}{\ensuremath{\mathrm{\,Ge\kern -0.1em V}}\xspace}
\newcommand{\mev}{\ensuremath{\mathrm{\,Me\kern -0.1em V}}\xspace}
\newcommand{\kev}{\ensuremath{\mathrm{\,ke\kern -0.1em V}}\xspace}
\newcommand{\ev}{\ensuremath{\mathrm{\,e\kern -0.1em V}}\xspace}
\newcommand{\gevc}{\ensuremath{{\mathrm{\,Ge\kern -0.1em V\!/}c}}\xspace}
\newcommand{\mevc}{\ensuremath{{\mathrm{\,Me\kern -0.1em V\!/}c}}\xspace}
\newcommand{\gevcc}{\ensuremath{{\mathrm{\,Ge\kern -0.1em V\!/}c^2}}\xspace}
\newcommand{\gevgevcccc}{\ensuremath{{\mathrm{\,Ge\kern -0.1em V^2\!/}c^4}}\xspace}
\newcommand{\mevcc}{\ensuremath{{\mathrm{\,Me\kern -0.1em V\!/}c^2}}\xspace}

\def\mum  {\ensuremath{{\,\upmu\rm m}}\xspace}

\def\mub{\ensuremath{{\rm \,\upmu b}}\xspace}

\def\invfb   {\ensuremath{\mbox{\,fb}^{-1}}\xspace}

\def\Xrad {\ensuremath{X_0}\xspace}
\def\NIL{\ensuremath{\lambda_{int}}\xspace}
\def\mip {MIP\xspace}

\def\gsim{{~\raise.15em\hbox{$>$}\kern-.85em
          \lower.35em\hbox{$\sim$}~}\xspace}
\def\lsim{{~\raise.15em\hbox{$<$}\kern-.85em
          \lower.35em\hbox{$\sim$}~}\xspace}

\def\pt         {\mbox{$p_{\rm T}$}\xspace}
\def\et         {\mbox{$E_{\rm T}$}\xspace}

\def\tell1  {TELL1\xspace}
\def\ukl1   {UKL1\xspace}

\setboolean{inbibliography}{false} 

\title{First Years of Running for the LHCb Calorimeter System}

\ShortTitle{First Years of Running for the LHCb Calorimeter System}

\author{\speaker{Pascal Perret}\thanks{On behalf of the LHCb Collaboration.}\\
        Clermont Universit\'e - CNRS/IN2P3 - Laboratoire de Physique Corpusculaire - BP 80026 - \\ F 63177 Aubi\`ere Cedex - France\\
        E-mail: \email{perret@in2p3.fr}}

\abstract{The calorimeter system of LHCb is subdivided into four sub-detectors which ensure its longitudinal segmentation: a Scintillator Pad Detector (SPD) followed by a Preshower (PS) and then an electromagnetic (ECAL) an hadronic (HCAL) calorimeter. After a description of these detectors, procedures developed to calibrate their response are discussed together with the degradation of response observed due to the harsh conditions (hard radiation and high current). 
First results on the calorimeter performance over the
two most productive years of physics data taking, 2011 and 2012 are given. 
}

\FullConference{Technology and Instrumentation in Particle Physics 2014,\\
		2-6 June, 2014\\
		Amsterdam, the Netherlands}

\begin{document}

\section{Introduction}
The \lhcb experiment~\cite{Alves:2008zz} is one of the four major experiments at the Large Hadron Collider (LHC) of CERN. 
The detector is a single-arm forward
spectrometer covering the \mbox{pseudorapidity} range $2<\eta <5$,
designed for the study of particles containing $b$ or $c$
quarks. It includes a high-precision tracking system
consisting of a silicon-strip vertex detector surrounding the $pp$
interaction, a large-area silicon-strip detector located
upstream of a dipole magnet with a bending power of about
$4{\rm\,Tm}$, and three stations of silicon-strip detectors and straw
drift tubes placed downstream.
The combined tracking system has a momentum resolution $\Delta p/p$ that varies from 0.4\% at 5\gevc to 0.6\% at 100\gevc, and an impact parameter (IP) resolution of 20\mum for tracks with high transverse \mbox{momentum (\pt)}. Charged hadrons are identified using two ring-imaging Cherenkov detectors (RICH). Photon, electron and hadron candidates are identified by a calorimeter system consisting of scintillating-pad and preshower detectors, an electromagnetic calorimeter and a hadronic calorimeter which is described in details in the next section. Muons are identified by a system composed of alternating layers of iron and multiwire proportional chambers. The trigger consists of a hardware stage, based on information from the calorimeter and muon systems, followed by a software stage which applies a full event reconstruction.

The LHC delivered luminosity at the LHCb interaction point of up to 4$\times 10^{32}$ cm$^{-2}$s$^{-1}$ for proton energies up to 8 \tev, which was equivalent
to the average number of interactions per beam crossing 1.8, while the design value for the LHCb luminosity was 2$\times 10^{32}$ cm$^{-2}$s$^{-1}$ resulting in an average number of interactions per beam crossing of 0.4, much smaller than LHC is capable to provide. 
In 2011 a luminosity leveling procedure was introduced.  By  adjusting the transverse overlap of the beams at LHCb the instantaneous luminosity could be kept
  stable to within about 5\% during a fill.
The two most productive years were 2011 and 2012, when 1.107 \invfb and 2.082 \invfb of integrated
luminosity have been recorded at beam energies 3.5 \tev and 4 \tev correspondingly.

The \bbbar cross-section measured by LHCb at 7 \tev constitutes 79 \mub~\cite{bbcrossect} in LHCb acceptance, implying $\sim$100 000 \bbbar pairs produced every second. In each p-p collision about 1500 charged particles are produced resulting in a loading challenging
radiation on detectors and readout electronics. In general, the apparatus has shown robust performance even in such
a challenging environment.

The performance of the calorimeters relies both on their time alignment~\cite{CaloTALP} and calibration~\cite{CaloPerformance}
which require different approaches for each sub-detector and are described in section~\ref{sec:MonitoringAndCalibration}, prior to detector ageing in section~\ref{sec:DetectorAgeing}. Some performances in terms of particle identification are illustrated in section~\ref{sec:Performance}. Additional results are available in Ref.~\cite{CaloPerformance} and references therein.

\section{Description of the LHCb Calorimeters}

The \lhcb calorimeter~\cite{CaloTDR} has two main goals: the first one is to provide information to the hardware trigger level and the second is to measure the neutral particle energy and to take part to particle identification for offline analysis. To reach them it is subdivided into four systems. From the interaction point particles encounter at 12,3~m the scintillating-pad detector (\spd), which is designed to identify charged particles, followed by a wall of lead and the preshower detector (\presh), then an electromagnetic calorimeter (\ecal) and a hadronic calorimeter (\hcal). This ensures the characterisation of the longitudinal profile of the showers, along the LHC beam axis.
All calorimeters use the same energy detection principle: scintillation
light is transmitted to a photomultiplier tube (PMT) by wavelength-shifting (WLS) 
fibres. 
The single fibres for the \spd/\presh cells are read out using multianode 
photomultiplier tubes (MAPMT), while the fibre bunches in the \ecal and \hcal
modules require individual phototubes. 
Optimal energy resolution requires the full containment of the showers from
high energy photons. For this reason, the thickness of \ecal was chosen to be 
25 radiation lengths ($\mathrm{X}_0$). The \presh is 10\% as thick as the \ecal.  
The \spd/\presh, \ecal and \hcal adopt a variable lateral segmentation into square cells since the hit density varies
by two orders of magnitude over the calorimeter surface. A
segmentation into three different sections has been chosen for the
\ecal and projectively for the \spd/\presh. Given the dimensions of the
hadronic showers, the \hcal is segmented into two zones with larger
cell sizes. The
outer dimensions of the \ecal match projectively those of the
tracking system, $\theta_x < 300 \ $mrad and $\theta_y < 250 \
$mrad while the square hole around the
beampipe limits the inner acceptance approximately to projective polar angles $\theta_{x,y} >
25 \ $mrad. The main parameters of the sub-detectors are summarised in Table~\ref{tab1}.

\begin{table}[htbp]
\begin{center}
\begin{tabular}{l|l|l|l}
sub-detector & \spd/\presh & \ecal & \hcal \\
\hline
number of channels & 2$\times$6016 & 6016 & 1488 \\
overall lateral & 6.2 m $\times$ 7.6 m & 6.3 m $\times$ 7.8 m & 6.8 m $\times$ 8.4 m \\
dimension in x, y & & & \\
cell size  (mm) Inner & 39.66 (\spd), 39,84 (\presh) & 40.4 & 131.3\\
\hphantom{cell size  (m)} Middle & 59.5  (\spd), 59.76 (\presh) & 60.6 & \\
\hphantom{cell size  (mm)} Outer & 119 (\spd), 119.5 (\presh) & 121.2 & 262.6\\
depth in z & 180 mm, & 835 mm, & 1655 mm, \\
           & 2 \Xrad, 0.1 \NIL & 25 \Xrad, 1.1 \NIL & 5.6 \NIL \\
light yield & $\sim$20 p.e./MIP & $\sim$3000 p.e./GeV &$\sim$105 p.e./GeV \\
dynamic range & 0 - 100 \mip   & 0 - 10 \gev \et & 0 - 20 \gev \et \\
            & 10 bits (\presh), 1 bit (\spd) & 12 bits & 12 bits \\
\end{tabular}
\caption{Main parameters of the \lhcb calorimeter sub-detectors}
\label{tab1}
\end{center}
\end{table}

The pad/preshower (\spd/\presh) detector uses scintillator pad readout
by WLS fibres that are coupled to MAPMT of 64 channels via
clear plastic fibres. 
The \spd/\presh detector consists of a 15 mm lead converter 2.5 \Xrad thick, that is
sandwiched between two almost identical planes of rectangular
scintillator pads of high granularity with a total of 12032 
detection channels. 

The shashlik calorimeter technology, i.e. a sampling
scintillator/lead structure readout by plastic WLS fibres, was
chosen for the electromagnetic calorimeter. 
The energy resolution of the \ecal modules was determined at
the test beam. The parametrisation 
$\sigma_E / E= $a$/ \sqrt{E} \oplus $b$ \oplus $c$/ E$ (E in GeV) 
is used, where a, b and c
stand for the stochastic, constant and noise terms respectively.
Depending on the type of module and test beam conditions the
stochastic and constant terms were measured to be $8.5\%<~ $a$ ~<
9.5\%$ and b $\sim$0.8\%. 

The hadron calorimeter (\hcal) is a sampling
device made from iron and scintillating tiles, as absorber and
active material respectively. Its thickness is limited to 5.6 nuclear interaction lengths (\NIL) due to space limitations and the limited performance required. The special feature of this sampling
structure is the orientation of the scintillating tiles that run
parallel to the beam axis. A resolution  of
$\sigma_E / E= (69\pm5)\%/ \sqrt{E} \oplus (9\pm 2)\%$ (E in GeV) was determined.

\section{Monitoring and calibration}
\label{sec:MonitoringAndCalibration}
The SPD calibration is performed by adjusting the discriminator
threshold for each channel to yield a pre-defined efficiency to
identify a charged track. The efficiency is computed from the ratio
between the number of hits with a track pointing to a given cell over
the total number of tracks pointing to it. 
The result of the efficiency as a function of the threshold can be compared with
the theoretical expectation given by the integral of the convolution of a Landau and a Poisson
distribution. The Landau distribution describes energy lost by charged particle traversing a
thin layer of material while the Poisson distribution describes the
fluctuations in the number
of photoelectrons generated at the photocathode. The resolution
in the electronics to set the threshold is 5\% of $E_{MIP}$ which directly sets the objective resolution
at the 5\% level.

The charge response of the 6016 PS channels is equalised according to the most probable energy
deposition of minimum ionising particles (MIPs). The goal is to match 10 ADC counts in the Front-End electronics (FE) to the MIPs peak value. It results in a 100 MIPs full dynamic. Discrimination of $e$ and $\gamma$'s at the hardware trigger is then achieved with a typical threshold of 5 - 10 MIP. By tuning the high voltage
(HV) of the MAPMT's one can adjust the gain of a group of 64 PS channels simultaneously.
The FE numeric gain with tunable values in the range [1; 2] further allows a fine adjustment of
the 64 channels amplitude response within a same board. This allow a calibration of the PS better than 5\%.

Both ECAL and HCAL are equipped with monitoring systems on the basis
of light emitting diodes (LEDs). Each photomultiplier is illuminated by light flashes of fixed
intensity, so the value of average PMT response can be used to follow up the behaviour of each
readout channel (mostly defined by the stability of the PMT gain). In order to eliminate the
effects, coming from possible variations of the intensity of LED flash, LEDs themselves are
monitored by means of very stable silicon PIN photodiodes.
The monitoring of ECAL and HCAL responses is performed online in parallel with data taking.
LEDs flash with the frequency of about 11 kHz synchronously with one of accelerator service
time slot not filled with protons and a subset at about 50 Hz is sent to a monitoring farm for analysis.
The LED system allow to follow fastly the ageing of the detector. It is also used to detect malfunctioning cells and bad cable connections.

The fine calibration of ECAL 
uses fits to the invariant mass distribution of $\piz\to\gamma\gamma$ decays. 
Photons
are defined from $3\times 3$ neutral clusters consistent with single
photon signals. The cell with the highest energy deposit is called the
seed. Each
\piz{} candidate mass is assigned to two cells, one for each photon
seed. When enough statistics is collected (for the most peripheral cells it amounts up to few hundred millions of interactions), the \piz{} mass
distribution is fitted for each cell, and the energy-scale of the cell
is corrected to center the mass peak onto the nominal \piz{} mass
  (135 MeV/$c^2$).
In addition, samples of pure electrons
selected from conversions are used to follow fast ageing of ECAL (Section~\ref{sec:DetectorAgeing})  by
studying the energy to momentum ratio (E/p) distributions.
This last method requires precise information from the tracking system. The final precision of the calibration is estimated
to be better than 2\%.

In addition to use of the LED system, more precise HCAL calibration is performed on the regular basis with a cesium
radioactive source calibration system. Two $~$10 mCi
$^{137}Cs$ sources (one per each detector half) are moved by a hydraulic system through
each scintillator tile. The PMT response is measured with a dedicated system of current
integrators. Relation between anode current and the measured particle energy is known
from test beam studies, which showed also that this method facilitates a cell-to-cell intercalibration
level 2-3\%. Such precision is more than enough to fulfill all HCAL objectives in LHCb. 

\section{Detector ageing}
\label{sec:DetectorAgeing}
Over the first years of almost continuous LHCb operation, signs of the degradation of detector response was observed in all four calorimeter sub-detectors. So far, two major reasons for it are identified. The first is the radiation damage of scintillator tiles and WLS fibers, which manifests itself in the areas with largest hit densities, especially
in the innermost regions around the beam pipe. This effect is proportional to the particle flux (charged and neutral particles). 
In the \ecal inner most region, up to $\sim$0.25 Mrad/year are observed.
For instance it was found that the precision of the \ecal monitoring based on LED suffers from the radiation degradation of transparency of polystyrene clear fibers used to transport the LED light to the ECAL photomultipliers. In order to improve the performance of the monitoring system, and especially in view of significant increase of \lhcb working luminosity foreseen after 2018, the present plastic fibers were replaced by radiation hard quartz fibers~\cite{yuri-calor2014}.
  
\begin{figure}[htbp]
  \begin{center}
    \resizebox{\textwidth}{!}{
      \includegraphics[height = 0.35\textwidth]{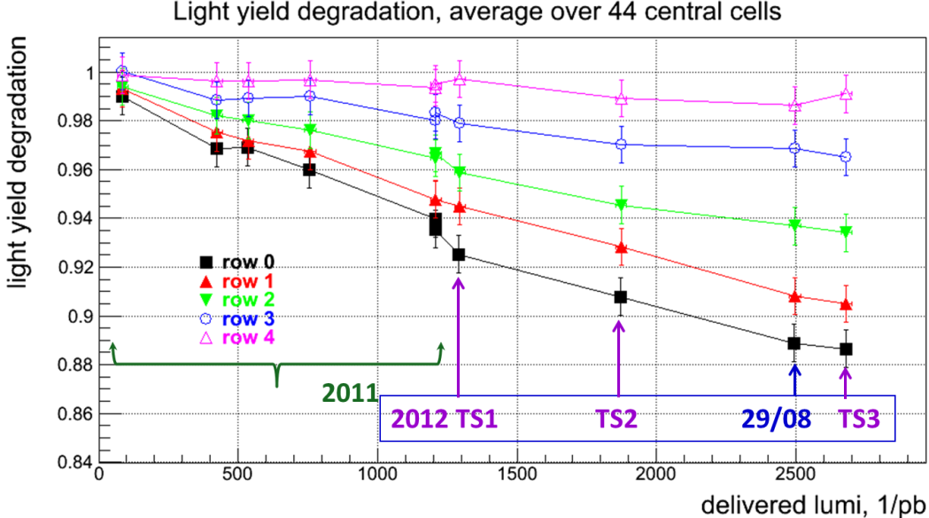}
      \includegraphics[height = 0.33\textwidth]{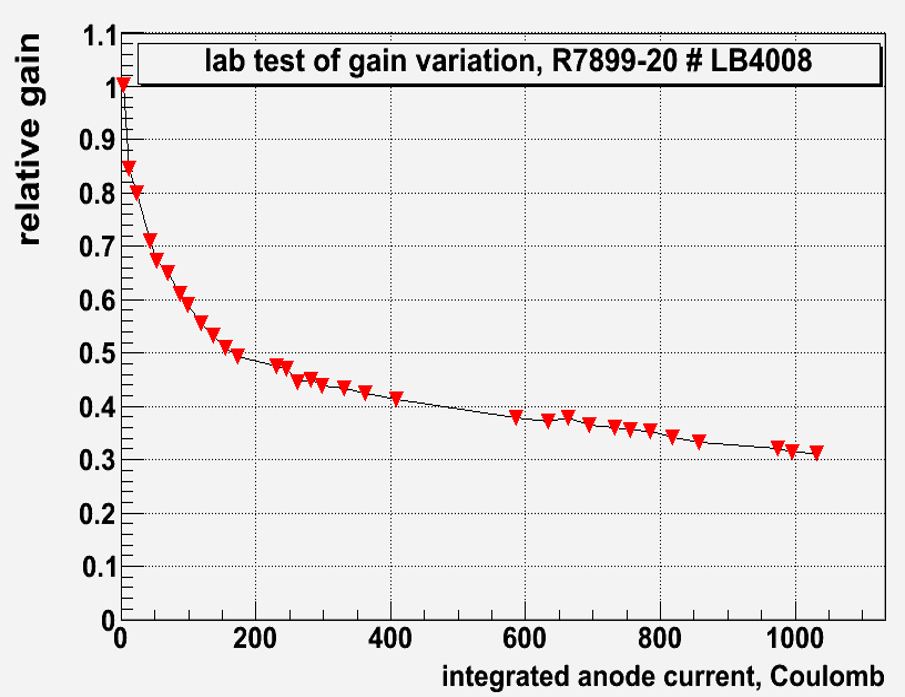}
    }
    \caption{Left: relative decrease of HCAL light yield for different layers of scintillator tiles, averaged over the 44 innermost cells, as a function of the delivered luminosity. Right: PMT gain degradation as a function of the integrated anode current measured at lab.\label{HCALplot} }
  \end{center}
\end{figure}

For now, the most extensive studies of its effects was done for HCAL, where light yields can be controlled directly with the $^{137}$Cs sources of the
calibration system. The average relative decrease in the light yield for 44 most central HCAL cells is shown in Fig.~\ref{HCALplot} (left) as a function of the luminosity, delivered by LHC over 2011 and 2012, separately for each layer of scintillator tiles. As expected, the most affected by radiation layer is
the frontal one, which corresponds to the shower maximum position, thanks to the additional 1.1 \NIL of ECAL.
The second factor is PMT sensitivity losses. The effect was studied in laboratory, the tested photomultiplier being kept under continuous illumination by short light pulses to emulate the \lhcb environment. Figure~\ref{HCALplot} (right) shows the relative change of PMT response with the integrated charge, which has passed through the device since the beginning of the test. The initial sharp decrease of the gain noticeably slows down after 150 coulombs, though it should be kept in mind, that this test
did not reproduce exactly the real operational conditions with periods of inactivity between fills, allowing some recuperation of PMT sensitivities.
This effect depends upon cell size and location. The actual amounts of integrated charge, obtained over 2011 and 2012, is estimated with
current integrators, which readings are continuously logged. In case of HCAL, the total charge, which  passed through the innermost photomultipliers during 2011, was on the level of 100 C.
Prior to the beginning of 2012 data taking, the gains of all HCAL PMTs were decreased by a factor of two in order to reduce the sensitivity losses and also to keep the PMT anode currents within the
recommended limits in the face of the increase of luminosity-setting. For ECAL, the maximal integrated charge was on the level of 20 C for 2011 and 37 C for 2012.

These are well known unavoidable effects which affect the trigger rate based on calorimeter information as well as detector performances. They required regular calibrations using the methods described in Sect.~\ref{sec:MonitoringAndCalibration} to restore uniform detector responses.
In addition the cumulative change in the response of the calorimeter sub-detectors could be followed with cell occupancies in minimum bias events, thanks to the luminosity leveling system providing relatively stable conditions over long periods of data taking. This method is very promising and preliminary studies have shown that a precision of $\sim$~1\% could be achieved after few hours of data taking. It will be implemented for an automatic and online calibration for the next data taking period (Run 2) to adjust PMT high voltage after each LHC fill of data taking to obtain a stable trigger. 

\section{Performance of the \lhcb calorimeters}
\label{sec:Performance}

\subsection{Electron identification}
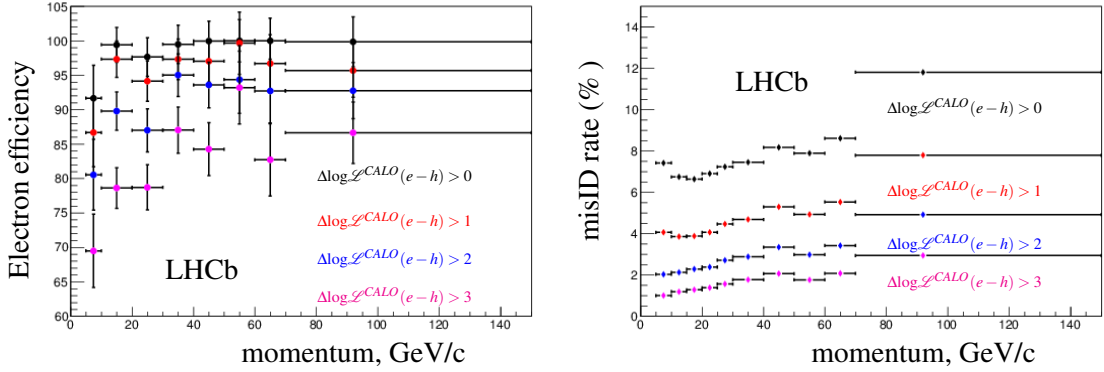
\begin{figure}[htbp]
\begin{minipage}[b]{0.5\linewidth}
\setlength{\unitlength}{1mm}
 \begin{picture}(0,50)
 \put(5,0){\includegraphics[width = 1.\textwidth]{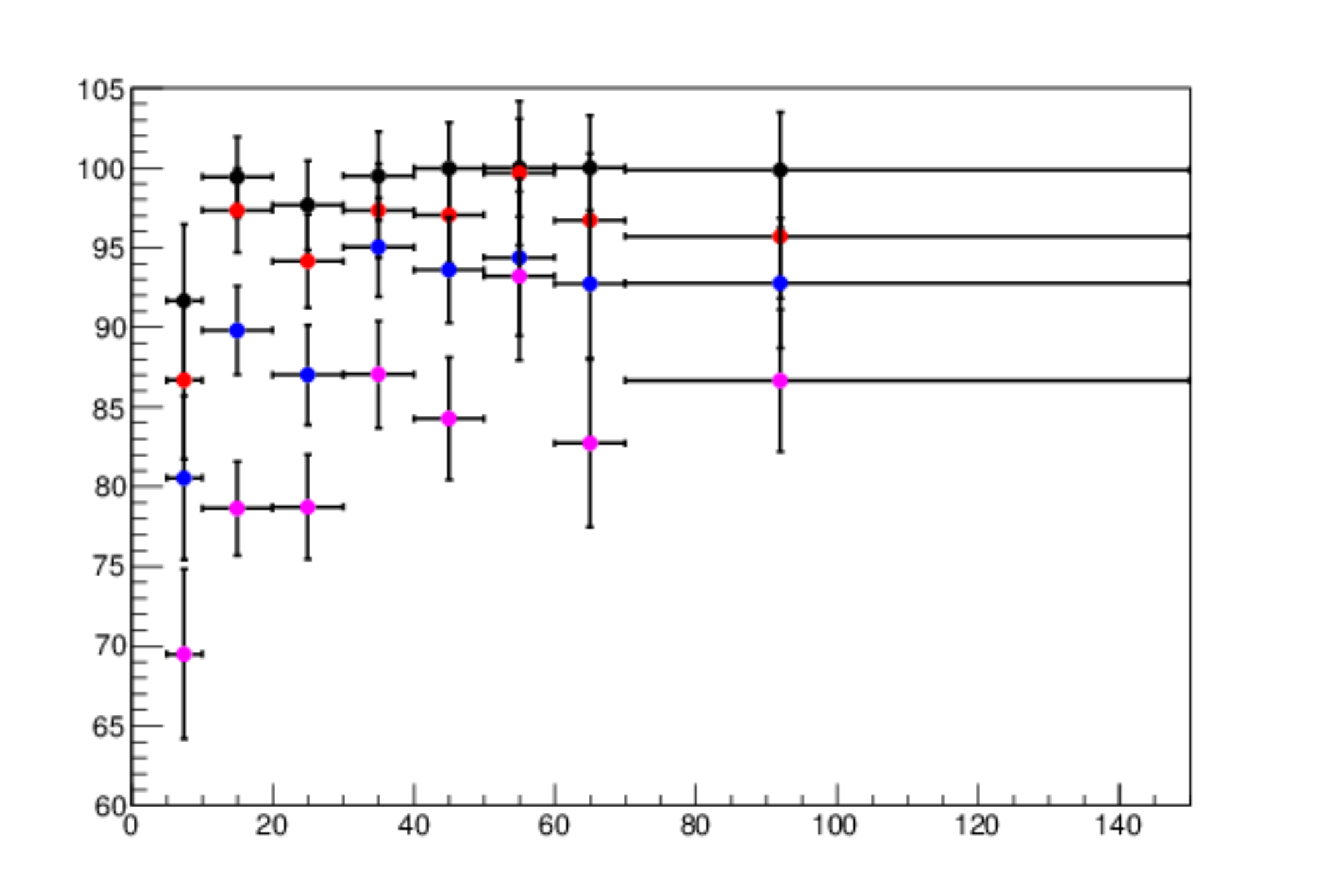}}
 \put(80,0){\includegraphics[width = 1.\textwidth]{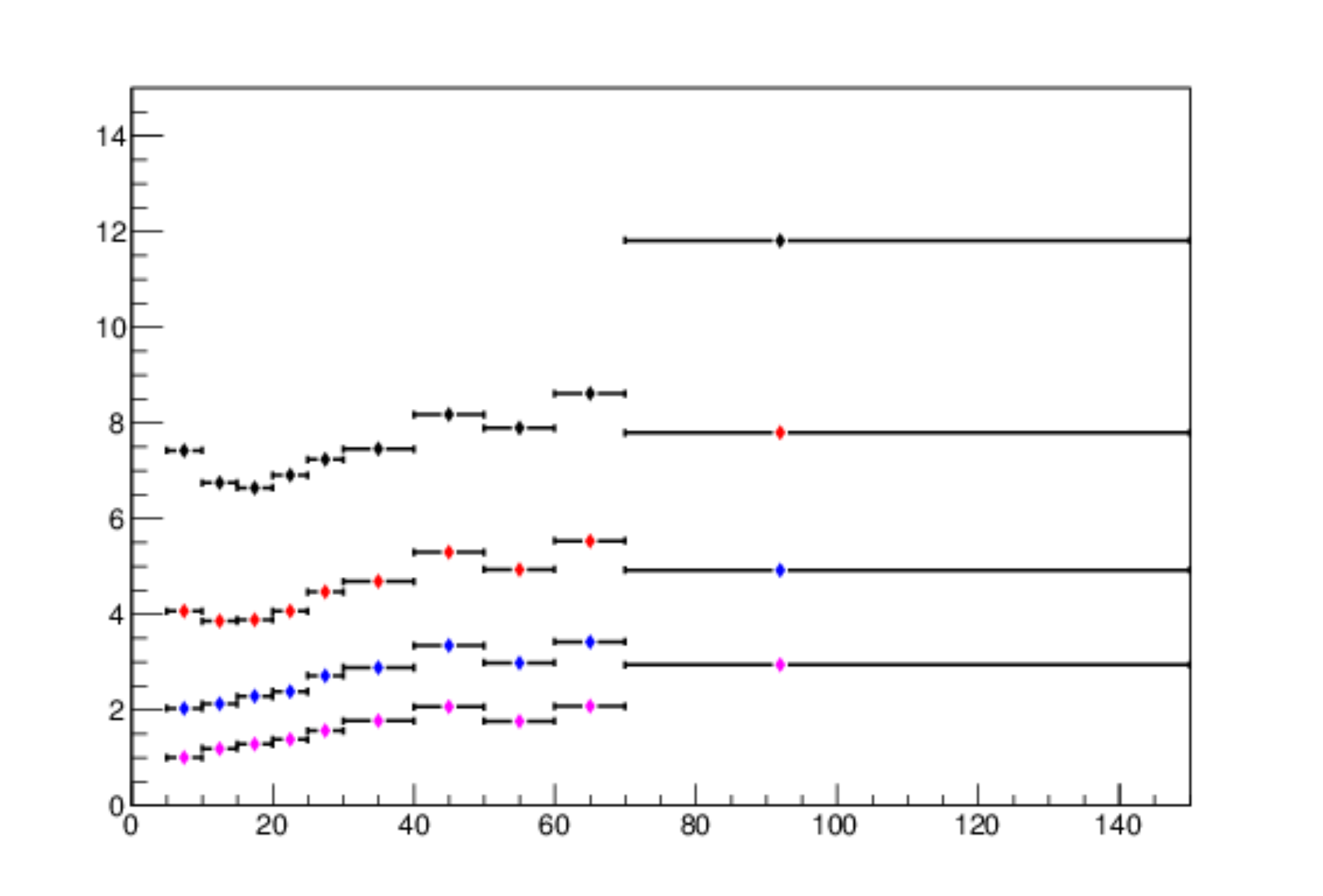}}
 \put(25,10) {LHCb}
 \put(100,35) {LHCb}
 \put(7,10) {\begin{rotate} {90} Electron efficiency\end{rotate}}
 \put(82,15) {\begin{rotate} {90} misID rate (\% ) \end{rotate}}
 \put(110,-1) {momentum, GeV/c }
 \put(35,-1) {momentum, GeV/c }
 \put(45,23) {\tiny {$\Delta$log${\mathcal L}^{CALO}(e-h) > 0$}}
 \put(45,17) {\tiny \textcolor{red}{$\Delta$log${\mathcal L}^{CALO}(e-h) > 1$}}
 \put(45,12) {\tiny \textcolor{blue}{$\Delta$log${\mathcal L}^{CALO}(e-h) > 2$}}
 \put(45,7) {\tiny \textcolor{magenta}{$\Delta$log${\mathcal L}^{CALO}(e-h) > 3$}}
 \put(120,32) {\tiny {$\Delta$log${\mathcal L}^{CALO}(e-h) > 0$}}
 \put(120,21) {\tiny \textcolor{red}{$\Delta$log${\mathcal L}^{CALO}(e-h) > 1$}}
 \put(120,14) {\tiny \textcolor{blue}{$\Delta$log${\mathcal L}^{CALO}(e-h) > 2$}}
 \put(120,9) {\tiny \textcolor{magenta}{$\Delta$log${\mathcal L}^{CALO}(e-h) > 3$}}
 \end{picture}
\end{minipage}
 \caption{\small Performance of the CALO system for electron ID with $\Delta {\rm log} {\mathcal L}^{CALO}(e-h)$ cuts : (right) efficiency as a function of momentum; (left)) misidentification rate as a function of momentum.
 \label{perf_calo} }
\end{figure}

The identification of electrons in the calorimeter system uses
information derived from the ECAL,
the PS  and the HCAL. The procedure
of combining these different sources of information
is based on signal and background likelihood distributions constructed for each sub-detector. In each case, reference 2-dimensional histograms correlating the energy measurement with the particle momentum are produced. 
One such histogram is filled for reconstructed electrons from gamma conversions and the other histogram for hadrons (pions and kaons from $D^0$- decays). After normalisation of the contents to unity in each slice of momentum, and taking the logarithm of the contents, the histograms represent the 2-dimensional log-likelihood  for electrons and hadrons. Their difference provides the 2-dimensional log-likelihoods of the signal and background hypotheses, which we denote $\Delta {\rm log} \mathcal L$. For a given track the electron identification is achieved by applying a cut on the value stored in the corresponding bin of the $\Delta {\rm log} \mathcal L$ histogram.
Since more than one estimator is available from LHCb calorimeter system, a combined one can be simply obtained by taking the product of the corresponding individual likelihoods:
\begin{eqnarray}
{\mathcal L}^{CALO} (e-h) = {\mathcal L}^{ECAL} (e-h) {\mathcal L}^{HCAL} (e-h) {\mathcal L}^{PS} (e-h)  \nonumber
\end{eqnarray}
The electron identification (ID) performance is evaluated using the data recorded in 2011. The available statistics allows the electron ID efficiency to be measured using a tag-and-probe method. The method is applied to $B^{\pm} \rightarrow J/\psi K^{\pm}$ candidates having $J/\psi \rightarrow e^+ e^-$, where one of the electrons is required to be identified by electron ID ($e_{tag}$) while the second electron is selected without using any information from the calorimeter system ($e_{probe}$). This second electron is used to estimate the efficiency of the electron ID and the performance of the electron $\Delta {\rm log} \mathcal L$ hypothesis test.
The efficiency $\epsilon_{CALO}$ and misidentification rate as a function of the $e_{probe}$ momentum is presented in Fig.~\ref{perf_calo} for several cuts on  $\Delta {\rm log} {\mathcal L}^{CALO}(e-h)$. By way of illustration of the typical performance, for electrons from $J/\psi \rightarrow e^+ e^-$ decay in $B^{\pm} \rightarrow J/\psi K^{\pm}$ events the average identification efficiency in the complete calorimeter system is (91.9$\pm$1.3)\% for a misidentification rate of (4.54$\pm$0.02)\% after requiring $\Delta {\rm log} {\mathcal L}^{CALO}(e-h)>$~2. In addition information from the RICH detectors can be used to improved these performances allowing to reach an average identification efficiency up to $97 \%$ for a misidentification rate lower than $2 \%$.

\subsection{Isolated photon and neutral pion reconstruction}
Energy deposits in ECAL cells are clusterised applying a 3x3 cell pattern around local maxima of energy deposition.
Consequently the centers of the reconstructed clusters are always separated at least by one cell.
If one cell is shared between several reconstructed clusters, the energy of the cell is
redistributed between the clusters proportionally to the total cluster energy.
The process is iterative and it converges rapidly due to relatively small ratio of the Moliere radius (3.5 cm) to the cell size. 
After the redistribution of energy of shared cells, the energy-weighted cluster moments up to the order 2 are evaluated to provide
the (hypotheses-independent) cluster parameters, namely, the total energy, the energy barycenter position and the two-dimensional energy spread matrix.
Neutral clusters are identified as those clusters that do not match to charged tracks.

The photon energy, $E_c$, is evaluated from the total cluster energy, $\epsilon_{ECAL}$, 
 and the reconstructed energy deposit in
the PS, $\epsilon_{PS}$, as follows: 
$
E_c = \alpha \times \epsilon_{ECAL} + \beta \times \epsilon_{PS}.
$
The parameters $\alpha$ and $\beta$ account for energy leakage.
The value of parameter $\alpha$ depends on the position of the cluster on the calorimeter surface while $\beta$ is estimated afterwards to give full account of energy samples.

\begin{figure}[htbp]
  \begin{center}
      \includegraphics[width = 0.4\textwidth]{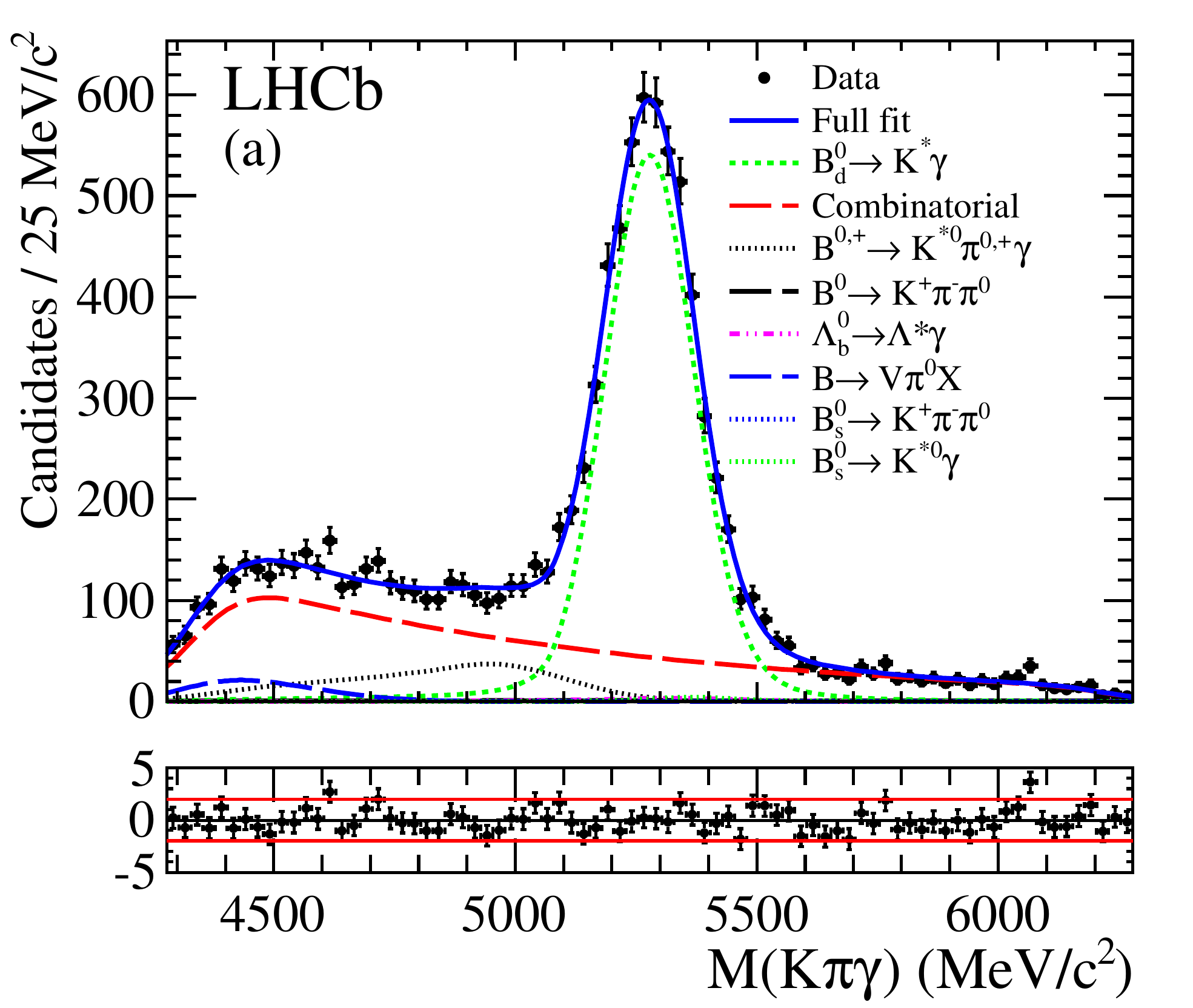}
			\includegraphics[width = 0.4\textwidth]{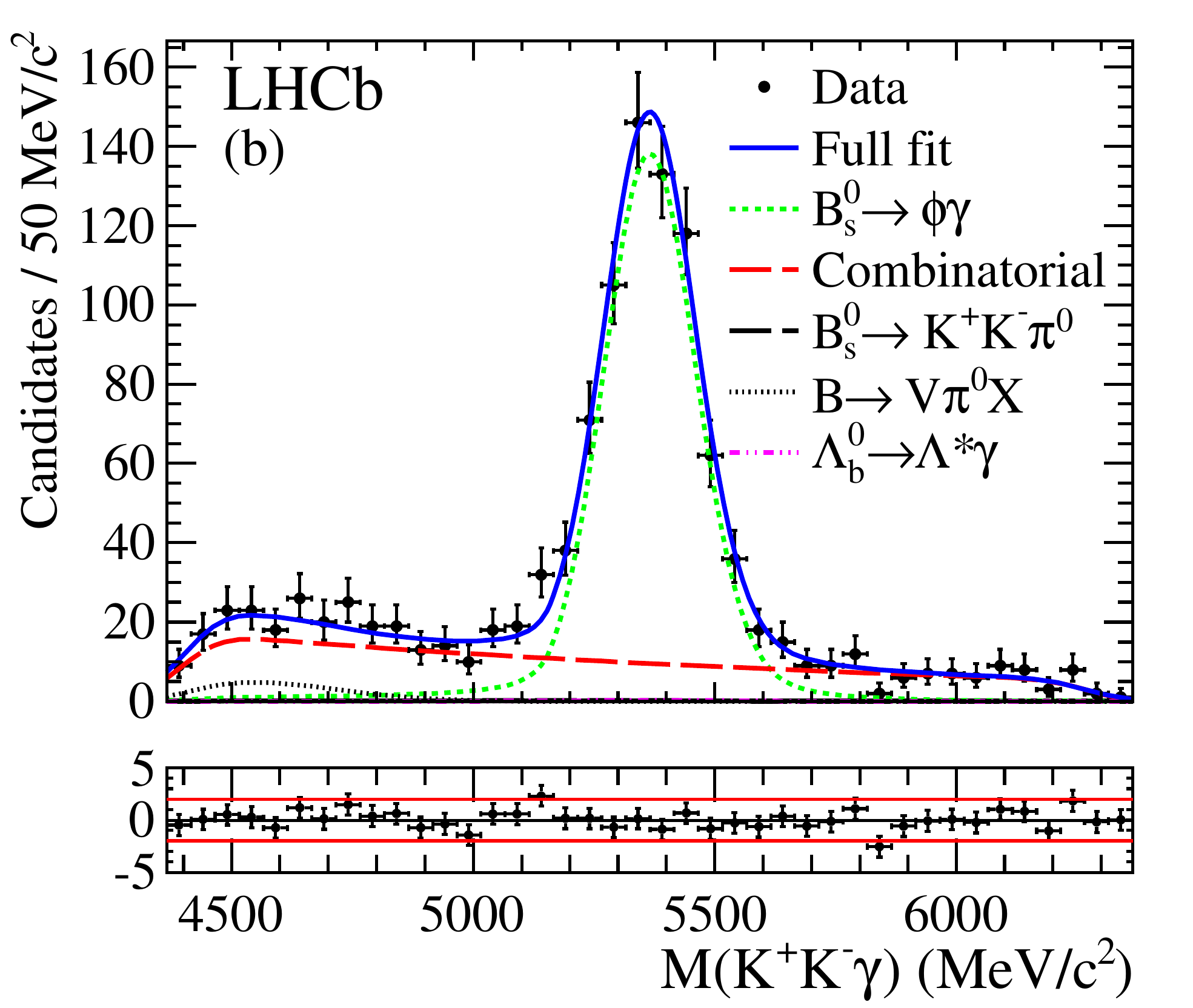}
    \caption{Mass distribution of the reconstructed $B^0\to K^{*0}(K^+ \pi^-)\gamma$ candidates (a) and $B^0_s\to \phi (K^+ K^-)\gamma$ candidates (b) obtained with an integrated luminosity of $\cal L$=1.0 $fb^{-1}$~\cite{BRrad}. The black points represent the data and the fit result is represented as a solid blue line. The signal (green dotted line) and the various background contaminations are shown. \label{Kstg} }
  \end{center}
\end{figure}

The performance of high energy photon reconstruction is illustrated by the reconstructed $B^0\to K^{*0}\gamma$ and $B^0\to \phi \gamma$ mass distribution~\cite{BRrad} shown in Fig.~\ref{Kstg}. The mass resolution obtained for this radiative decay, $\sim$92 $MeV/c^2$, is dominated by the ECAL energy resolution. A comparison of the data to simulated samples shows that this corresponds to an intrinsic resolution in the photon energy of around 2\%~\cite{CaloPerformance}.

Neutral pions with low transverse energy are mostly reconstructed as a resolved pair of well separated
photons.  A mass resolution of 8{\mevcc} is obtained for such neutral pions.
However, due to the finite ECAL granularity, photon pairs from the
decay of sufficiently high momentum \piz{} cannot be resolved as
individual clusters. This essentially holds for all \piz{} decays with
transverse momentum above 2 {\gevc}. To reconstruct such \piz{}
candidates a procedure has been design to identify merged clusters. 
The algorithm consists of splitting each single ECAL clusters into
two $3\times3 $ subclusters built around the two highest deposits of the original cluster.
The energy of the common cells is then distributed over the two assumed subclusters according to an iterative procedure based on the expected transversal shape of photon showers.
The division of the energy depends on the barycenter position of each subcluster. Since the latter is a function of the energy distribution, this procedure requires a few iterations, but it is found to converge quickly. A $\piz\to\gamma\gamma$ invariant mass resolution of $\sim$30{\mevcc} is obtained for these `merged' $\piz$ candidates while it is closed to $\sim$20{\mevcc} for `resolved' $\piz$.

\setboolean{inbibliography}{true}

\end{document}